# Nanoscale imaging of unusual photo-acoustic waves in thin flake VTe$_2$

A. Nakamura[1*†], T. Shimojima[1†], Y. Chiashi[2], M. Kamitani[1], H. Sakai[3], S. Ishiwata[2,4], H. Li[2], K. Ishizaka[1,2]

[1]*RIKEN Center for Emergent Matter Science, Wako, Saitama 351-0198, Japan.*
[2]*Quantum-Phase Electronics Center and Department of Applied Physics, The University of Tokyo, Hongo, Tokyo 113-8656, Japan.*
[3]*Department of Physics, Osaka University, Toyonaka, Osaka 560-0043, Japan.*
[4]*Division of Materials Physics, Graduate School of Engineering Science, Osaka University, Toyonaka, Osaka 560-8531, Japan.*

Controlling acoustic phonons, the carriers of sound and heat, has been attracting great attention toward the manipulation of sonic and thermal properties in nanometric devices. In particular, the photo-acoustic effect using ultrafast optical pulses has a promising potential to optically manipulate phonons in picoseconds time regime. However, its mechanism has been so far mostly based on the commonplace thermoelastic expansion in isotropic media, limiting the spectrum of potential applications. We investigate a conceptually new mechanism of photo-acoustic effect involving the structural instability, by utilizing a transition-metal dichalcogenide VTe$_2$ with the ribbon-type charge-density-wave (CDW). Ultrafast electron microscope imaging and diffraction measurements reveal the generation and propagation of unusual acoustic waves in the nanometric thin plate associated with the optically induced instantaneous charge-density-wave dissolution. Our results highlight the capability of photo-induced structural instability as a source of coherent acoustic waves.

Photoacoustic effect, one major interaction between light and phonons, has been long studied since its discovery in 1880's[1,2], where modulated light produces acoustic waves. The recent development of femtosecond laser technology further provided a new route, for the coherent control of acoustic phonons in picosecond ($10^{-12}$ s) time regime. When a material is irradiated by an ultrashort optical pulse, the photoexcited near-surface area of the sample is instantaneously deformed. Generated pulsed strains, referred to as coherent acoustic phonons (CAP), propagate in materials with sound velocity accompanying the coherent atomic motions[3,4]. When the size of medium is reduced to nanometric, it becomes comparable to the length-scale of the strain pulse (*i.e.* the optical penetration depth), and the acoustic resonance occurs in GHz–THz regime. Thus, the photoacoustic effect combined with nanofabrication has a large potential for realizing the ultrafast and coherent control of phonons and related properties. Nevertheless, past investigations are mostly originated from the ubiquitous photothermal expansion with predominantly isotropic deformations[5] (see Fig. 1a) *i.e.* longitudinal acoustic displacements[6], with a very few exceptional efforts utilizing piezoelectricity[7–9].

Here we investigate a new CAP generation mechanism by focusing on the structural instability of materials. In general, a structural phase transition modifies the lattice parameters, and in some cases, leads to an anisotropic lattice deformation (see Fig. 1b). This is strongly associated with the specific acoustic phonons and provides us with a chance to selectively control the strains. In this study, we focus on charge-density-wave (CDW) transition-metal dichalcogenide VTe$_2$, which shows an anisotropic trigonal-monoclinic structural phase transition reflecting the characteristic shear deformation of VTe$_2$ layer stacking. In general, CDW materials show the structural instabilities coupled to electronic states via the strong electron-phonon interaction. Thus, the photo-induced structural phase transitions of sub-ps timescales have been indeed reported in many CDW systems[10–13]. Figures 1c and 1d show the crystal structures[14,15] and the electron diffraction patterns of VTe$_2$. Above the structural phase transition temperature $T_s$ (~ 480 K), VTe$_2$ has a simple CdI$_2$-type (1$T$) structure with a space group of P$\bar{3}$m1. Below $T_s$, the vanadium-vanadium bond structure reflecting the CDW[16] is formed. Correspondingly, some additional spots originating from the superstructure, referred to as CDW peaks, are observed in the electron diffraction pattern. The symmetry changes to monoclinic C2/m (1$T''$ phase[17]), and consequently, the shear deformation of the crystal characterized by the angle $\beta > 90°$ in Fig. 1d occurs. Here we report the spatio-temporal evolution of CAPs in VTe$_2$ thin flakes by ultrafast electron microscopy (UEM). As described in previous studies[18–20], the CAPs propagating along perpendicular to the flake (*i.e.* the direction of the phonon wavevector $\boldsymbol{k} \parallel \boldsymbol{e}_z$, where $\boldsymbol{e}_z$ is the unit vector along the $z$-axis in Fig. 1c) can be detected by the time-resolved diffraction measurement. As photo-generated CAPs cause deformation of the crystal plane, the intensity of diffracted electrons temporally oscillates through the change in diffraction (Bragg) condition[18,19]. Correspondingly, the acoustic phonons propagating laterally along the flake (*i.e.* $\boldsymbol{k} \perp \boldsymbol{e}_z$) are further observable through the bright-field imaging of UEM [21,22].

UEM measurements are performed using a Technai Femto (FEI) with an acceleration voltage of 200 kV (See crystalline VTe$_2$ with the thickness approximately 75 nm are prepared using an ultramicrotome, and mounted on the copper mesh grid of 150 μm for transmission electron microscope measurements.

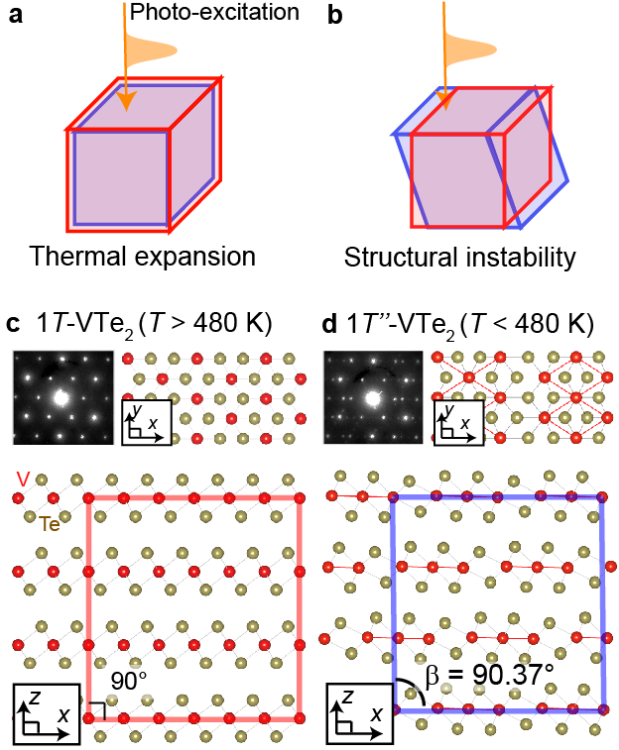

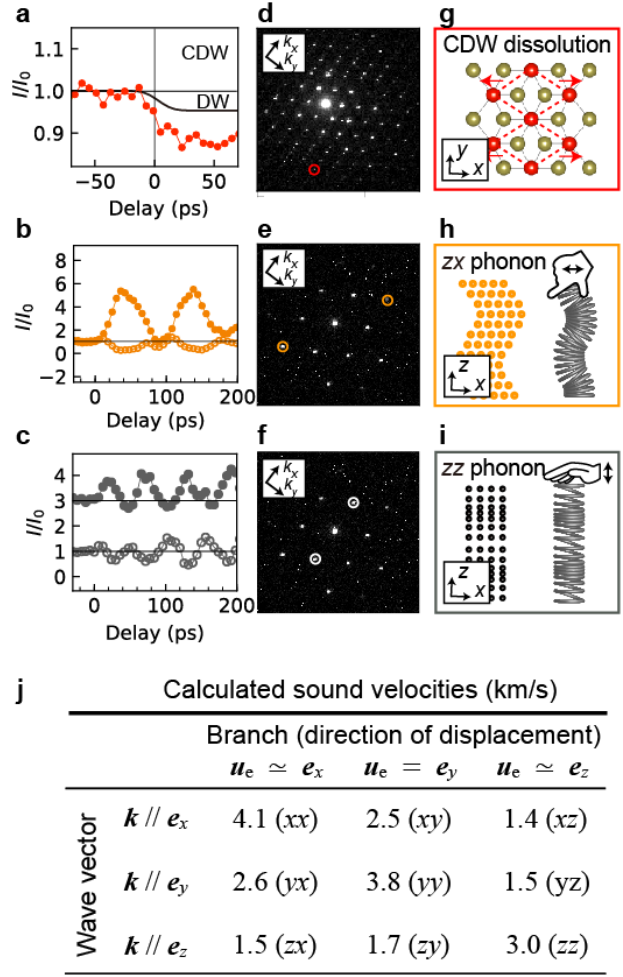

**Figure 1.** Schematic of photo-acoustic effect and VTe$_2$ structural properties. (a, b) Schematic of the photo-induced lattice deformation induced by the conventional photothermal expansion and the structural instability of crystals, respectively. (c, d) Crystal structures and static electron diffraction patterns of 1$T$- and 1$T''$(CDW)-VTe$_2$, respectively. The monoclinic angle $\beta$ changes from 90° to 90.37° in 1$T''$ phase. $x$, $y$, and $z$ axes denote Cartesian coordinates (see Fig. S3).

**Figure 2.** Ultrafast dissolution of CDW and subsequent acoustic phonon generations. (a) Time evolution of $\overline{22/3}\,\overline{2}\,1/3$ diffraction intensity. The solid black curve shows calculated contribution of Debye–Waller (DW) factor. (b) The solid and open markers show $I(t)/I_0$ of $3\,\overline{1}\,0$ and $\overline{3}\,1\,0$, respectively. (c) $I(t)/I_0$ of $2\,0\,0$ (solid markers) and $\overline{2}\,0\,0$ (open markers). $I/I_0$ of $2\,0\,0$ is offset by 2. All data in (a-c) were recorded with a pump fluence of 0.18 mJ/cm$^2$. (d-f) Corresponding diffraction images for the data in (a-c). The solid circles show the diffraction spots used for respective analysis. The fields of view used to obtain the diffraction images are shown in Fig. S5. (g) Schematic of the CDW dissolution. (h, i) Schematics of $zx$ and $zz$ phonons, respectively. (j) Calculated bulk sound velocities of 1$T''$-VTe$_2$. $k$ and $u_e$ denote the wave vector and normalized atomic displacement vector of the phonon, respectively. Details of the calculations are shown in Supplemental Materials (Section 7).

Fig. S2). A 40-μm carbon-coated LaB$_6$ tip (Applied Physics Technologies) is used as a photocathode. The 150-μm condenser aperture is used for all measurements. For the bright-field measurements, we use a 20-μm objective aperture. As a photon source, we use an Yb:KGd(WO$_4$)$_2$ solid-state laser (Light Conversion, PHAROS). The generated laser beam is split into pump and probe beams. The fundamental 1030-nm pump beam with a pulse duration of 290 fs passes through a delay line and is used for exciting the sample. The rest of the beam passes through two $\beta$-Ba$_2$B$_2$O$_4$ (BBO) crystals for fourth-harmonic generation (257-nm), and is focused on the photocathode in the microscope. The repetition rate of the measurement was set to 10 kHz. The best time resolution is ~2 ps, nevertheless, it was set to ~15 ps for the data acquisition in Figs. 2 – 4 to keep the good S/N ratio throughout the whole measurement. The single crystals of VTe$_2$ are synthesized by the chemical vapor transport method with TeCl$_4$ as a transport agent. The temperatures at the source and growth zones are set to 600 and 550 °C, respectively. The free-standing thin flakes of single-

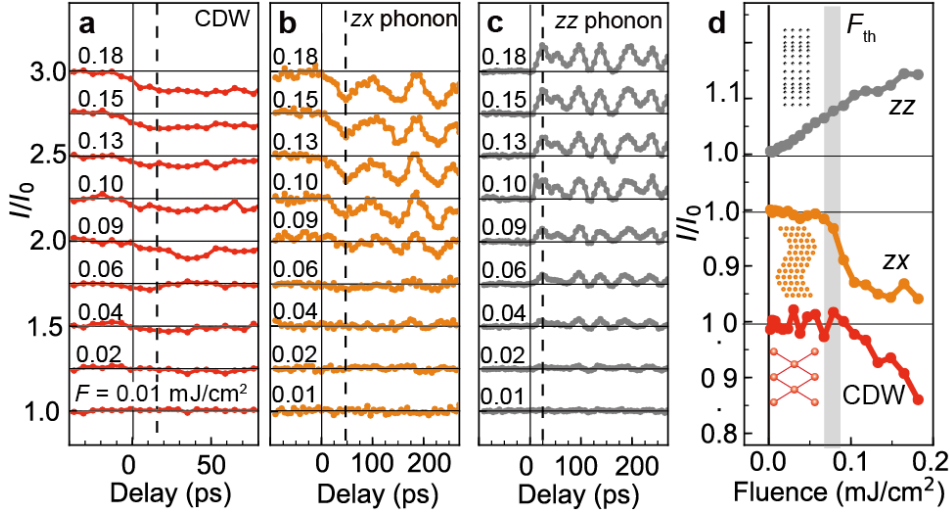

**Figure 3.** Pump-fluence dependence of CDW dissolution and *zx*, *zz* coherent acoustic phonons. (a-c) Fluence (*F*)-dependent $I/I_0$ curves for the $\overline{22/3}\ \overline{2}\ \overline{1/3}$, $\overline{19/3}\ \overline{1}\ \overline{1/3}$, and 3 1 0 diffraction spots, respectively. The data are shown with vertical offsets for clarity. $I/I_0$ curves in (a, b, c) mostly reflect the CDW dissolution, *zx*, and *zz* phonons, respectively. (d) Fluence dependences of $I/I_0$ for $\overline{22/3}\ \overline{2}\ \overline{1/3}$ (bottom), $\overline{19/3}\ \overline{1}\ \overline{1/3}$ (middle), and 3 1 0 (top) at fixed delay times, respectively, indicated by the broken black lines in (a-c). The threshold fluence $F_{th}$ (0.08 mJ/cm²) is indicated by the gray shaded line.

First we investigate the photo-induced CDW dissolution in VTe$_2$ by analyzing the time (*t*)-dependent CDW peak intensity $I(t)$. We use the practical lattice vectors $\overline{a}, \overline{b},$ and $\overline{c}$, to describe the diffraction indices (see Fig. S3). The $I(t)/I_0 = I(t)/I(t<0)$ curve for the $\overline{22/3}\ \overline{2}\ \overline{1/3}$ CDW peak (see Figs. 2a and 2d) shows suppression by 12% within the first 15 ps, *i.e.* the time resolution of this measurement. The timescale of the CDW dissolution is indeed faster than 2 ps as confirmed by a measurement of higher *t* resolution (see Fig. S6). Here, the maximum heating of the lattice temperature is estimated to be ~24 K at the present pump fluence (0.18 mJ/cm²). The expected intensity reduction by this heating, *i.e.* the Debye–Waller factor, is overlaid in Fig. 2a (see Section 3 in Supplemental Materials). The $\overline{22/3}\ \overline{2}\ \overline{1/3}$ intensity decreases significantly compared to the Debye–Waller factor, thus suggesting the suppression of the superlattice peak due to the photoinduced CDW dissolution in VTe$_2$ (Fig. 2g). We note that the previous ultrafast diffraction studies on transition metal dichalcogenides, e.g. TaS$_2$, TiSe$_2$, TaSe$_2$ [10–13], reported similar CDW melting on optical excitation.

The transient diffraction intensities in a longer timescale further reveal the subsequent acoustic phonon dynamics. Figs. 2b and 2c clearly show the temporal oscillations of diffraction intensities with time periods of $T \simeq 91 \pm 6$ ps and $T \simeq 50 \pm 5$ ps, respectively. The oscillations of $h\ k\ 0$ and $\overline{h}\ \overline{k}\ 0$ intensities (see Figs. 2e and 2f) are clearly opposite in sign. This asymmetric behavior cannot be explained by the change of crystal structure factor, and indicates the shear lattice deformation as discussed previously[19,23]. We further note that these CAPs are observed through the slight change of the diffraction angle, and their appearance strongly depends on the experimental setup such as sample tilting, rather than the Miller index itself. In the previous ultrafast electron diffraction/microscopy studies[19–23], such diffraction intensity oscillations have been attributed to the standing-wave of CAPs satisfying the condition: $T = 2d/v$, where $d$ and $v$ denote the sample thickness and the sound velocity of the longitudinal acoustic mode, respectively. Considering the present case, $T \simeq 91$ ps and 50 ps with $d = 75$ nm provide the sound velocities of 1.6 km/s and 3.0 km/s, respectively. For comparison, the sound velocities of VTe$_2$ obtained by first-principles calculation are shown in Fig. 2j. The *ij* phonon indicates those propagating along $\boldsymbol{k} \parallel \boldsymbol{e}_i$ with normalized atomic displacement along $\boldsymbol{u}_e \simeq \boldsymbol{e}_j$. Here we can assign the oscillations of $T \simeq 91$ ps and 50 ps to the standing-waves of longitudinal (*zz*, Fig. 2i) and transverse (*zx*, Fig. 2h) acoustic phonons, respectively (see Section 7 in Supplemental Materials for the calculation of sound velocities and branch assignment). We note that the density-functional calculations with generalized-gradient approximation tend to underestimate the elastic constants and sound velocities[24], which may partly explain the small mismatch of the observed and calculated sound velocities. As acoustic phonons with transverse displacement are hardly excited by the conventional thermoelastic scenario[12], the emergence of *zx* mode implies a new mechanism of photoacoustic effect related to the peculiar structural instability of VTe$_2$.

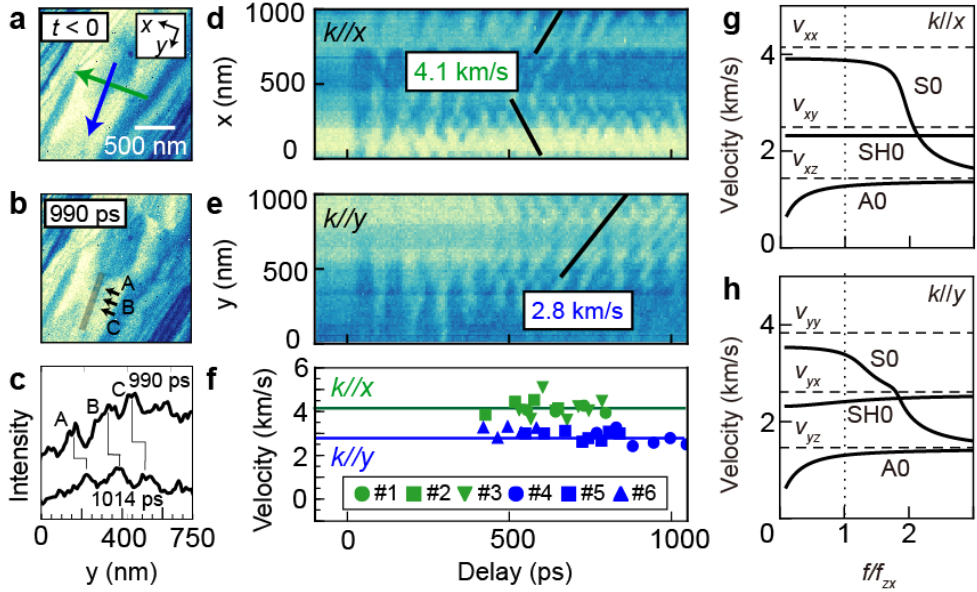

**Figure 4.** Lateral propagation of acoustic phonons in VTe$_2$. (a) Bright-field image of VTe$_2$ before photoexcitation. The solid green and blue arrows indicate the region where the analysis in (d) and (e) was performed. (b) Bright-field image at 990 ps after photoexcitation ($F = 0.18$ mJ/cm$^2$). The wave fronts of CAPs are indicated by black arrows (A, B, and C). (c) Line profiles of bright-field intensities along the gray shaded line in (b) recorded at 990 and 1014 ps. Three wave fronts, A, B, and C show the ~60 nm shift. (d, e) Dynamics of the bright-field contrast along $x$ and $y$, respectively, indicated by the solid green and blue arrows in (a). The solid black lines indicate the propagation of the wave fronts. (f) The sound velocities along $x$ and $y$ at different sample positions (#1–6). The green (#1–3) and blue (#4–6) markers represent the sound velocities along $x$ and $y$ estimated from Fig. S1, respectively. Solid green (blue) line represents the estimated sound velocity of 4.1 km/s (2.8 km/s). (g, h) Calculated phase velocities of zeroth-order symmetric (S0), asymmetric (A0), and shear horizontal (SH0) waves propagating along $x$ and $y$ direction, respectively. The horizontal axes are normalized by the resonant frequency of $zx$ mode ($f_{zx} = v_{zx}/2d$). These dispersion curves are calculated by the partial wave technique described in the literature[25]. $v_{ij}$ denotes the bulk sound velocity of the $ij$ phonon as shown in Fig. 2j.

To confirm the origin of the unusual $zx$ phonon generation, we perform the similar measurements with various pump fluences ($F$). Figure 3a shows the $I(t)/I_0$ curves of $\overline{22/3}\ \overline{2}\ \overline{1/3}$ CDW peak for respective $F$. The $F$-dependent $I/I_0$ recorded at $t = 15$ ps shown in Fig. 3d indicates that the CDW begins to dissolve above the threshold of $F_{\text{th}} = 0.08$ mJ/cm$^2$. This result is reminiscent of the fluence threshold observed in other systems with photo-induced phase transitions[11]. Similarly, as shown in Fig. 3b, the $zx$ phonon oscillation with $T \simeq 91$ ps begins to evolve at $F > F_{\text{th}}$. The $F$-dependence in Fig. 3d indicates that CDW dissolution and $zx$ phonon emerge above the identical threshold fluence $F_{\text{th}}$. It strongly suggests that the unusual $zx$ phonon generation is caused by the instantaneous CDW dissolution occurring in $t < 2$ ps. Here we stress that the CDW in VTe$_2$ couples to the shear (monoclinic) deformation as shown in Fig. 1. Conversely, the $F$-dependence of the longitudinal $zz$ phonon amplitude in Figs. 3c and 3d shows a rather monotonic increase, indicating that $zz$ phonons are predominantly generated by the conventional instantaneous photothermal expansion.

Finally, we reveal the lateral propagations of these CAPs along the thin flake, where the acoustic strains travel in the plate-wave form[25]. In the $t$-dependent bright-field images in Movies 1–4, the in-plane CAP propagations are observed at various sample positions. To show this, the bright-field images before ($t < 0$) and after photoexcitation ($t = 990$ ps) are presented in Figs. 4a and 4b, respectively. The propagating wave fronts of CAPs, indicated by black arrows in Fig. 4b, show the ~60 nm shifts along $y$ in the 24 ps time interval as shown in Fig. 4c. We note that the dynamics of bright-field images are mostly observed only at $F > F_{\text{th}}$, as mentioned in the Section 8 in Supplemental Materials (Fig. S7), indicating that they are predominantly triggered by the CDW dissolution. Figs. 4d and 4e show the $t$-dependent cut of the bright-field image along $x$ and $y$, as indicated by the green and blue arrows in Fig. 4a, respectively. Immediately after the photoexcitation, the bright-field intensities are dominated by the temporal oscillation of $T \simeq 91$ ps, the standing-wave of $zx$ phonon discussed in Fig. 2b. At $t > 500$ ps, the lateral propagations of CAPs along $x$ and $y$ are observed, as indicated by the solid black lines. The velocities estimated by the gradient of these lines are approximately 4.1 and 2.8 km/s, respectively, for $x$ and $y$ directions. These anisotropic velocities are confirmed in a wide $t$ range and also at different sample positions (see Fig. 4f). We note that the plate-waves that can propagate in this case are the zeroth order symmetric and asymmetric Lamb waves (S0, A0) and the zeroth order shear horizontal (SH0) wave[25]. Figs. 4g and 4h show the velocities of these modes calculated by the partial wave technique[25], for the present VTe$_2$ thin-flake. The velocities of S0, SH0, and A0 waves at the resonant frequency of $zx$ standing wave ($f_{zx} = v_{zx}/2d$) for $\boldsymbol{k} \parallel \boldsymbol{e}_x$ are estimated to be 93%, 92%, and 88% of bulk sound velocities $v_{xx}$, $v_{xy}$, and $v_{xz}$, respectively, indicating that their properties are fairly close to the corresponding bulk sound waves. Similarly, the velocities of S0, SH0, and A0 for $\boldsymbol{k} \parallel \boldsymbol{e}_y$ are estimated to be 89%, 92%, and 88% of $v_{yy}$, $v_{yx}$, and $v_{yz}$, respectively. Considering that the calculations tend to include certain underestimation as already mentioned [24], the observed CAPs with the sound velocities of 4.1 km/s ($\boldsymbol{k} \parallel \boldsymbol{e}_x$) and 2.8 km/s ($\boldsymbol{k} \parallel \boldsymbol{e}_y$) should be S0 ($xx$) and SH0 ($yx$) phonons, respectively, both with dominant atomic displacements along $x$. Appearing of the $yx$ mode in the thin flake (*i.e.* shear horizontal wave) is very unusual, considering that the photoacoustic phonons generated through the commonplace processes of thermoelastic and defect-mediated mechanisms are generally dominated by the longitudinal displacements[22,26]. Our present result thus strongly suggests the potential of utilizing the built-in structural instability for exciting specific acoustic waves.

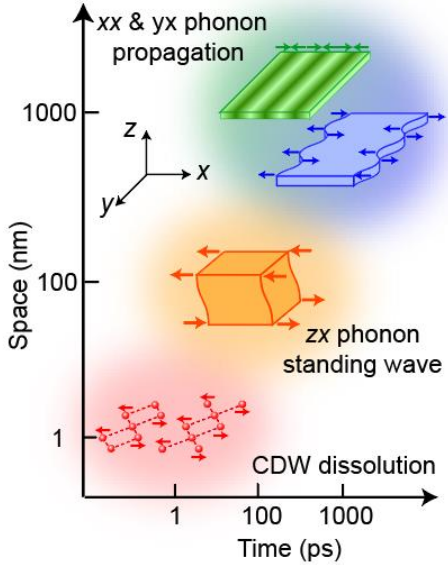

**Figure 5.** Temporal and spatial evolution of unusual photoacoustic process in VTe$_2$. Three phonon modes with $x$-axis atomic displacements ($xx$, $yx$, $zx$) are generated through the shear structural instability, triggered by the CDW dissolution.

The spatio-temporal evolution of observed phononic phenomena in thin-flake VTe$_2$ is summarized in Fig. 5. The ultrafast CDW dissolution in sub-nm and sub-ps microscopic regime triggers the peculiar CAPs dynamics on sub-μm and sub-ns scales through the shear instability of the monoclinic crystal coupled to CDW. These CAPs are characterized by the $x$-atomic displacement as represented by the observed $xx$, $yx$, and $zx$ phonons. This study highlights the large potential of the photo-induced phase transitions as a new generation mechanism of coherent acoustic waves, which will bring the future phononic and optoacoustic applications in the ultrafast time regime.

## ACKNOWLEDGEMENTS

We thank T. Nemoto for thin-flake preparation and X.Z. Yu for discussion.